\documentclass{eas}
\usepackage{graphicx}
\usepackage{amsmath}
\usepackage{graphicx}
\usepackage{txfonts}
\usepackage{subfig}
\usepackage{booktabs}
\usepackage{upgreek}
\usepackage{setspace}
\usepackage{gensymb}
\usepackage{comment}
\usepackage{multirow}
\usepackage{changepage}
\usepackage{float}
\usepackage{appendix}
\usepackage{tabularx}
\usepackage{epsfig}

\begin{document}
\title{The GK Persei nova shell and its $\lq$jet-like' feature} 
\author{E. Harvey}\address{Centre for Astronomy, School of Physics, National University of Ireland Galway, Galway, Ireland}
\author{M.P. Redman$^1$}
\author{P. Boumis}\address{National Observatory Athens, IAASARS, I. Metara \& V. Pavlou, Penteli, GR-15236 Athens, Greece}
\author{S. Akras}\address{Observat\'orio do Valongo, Universidade Federal do Rio de Janeiro, Ladeira Pedro Antonio 43 20080-090 Rio de Janeiro, Brazil}
\begin{abstract}
GK Persei (1901, the $\lq$Firework Nebula') is an old but bright nova remnant that offers a chance to probe the physics and kinematics of nova shells. The kinematics in new and archival longslit optical echelle spectra were analysed using the {\sc shape} software. New imaging from the Aristarchos telescope continues to track the proper motion, extinction and structural evolution of the knots, which have been observed intermittently over several decades. We present for the first time, kinematical constraints on a large faint $\lq$jet' feature, that was previously detected beyond the shell boundary. These observational constraints allow for the generation of models for individual knots, interactions within knot complexes, and the $\lq$jet' feature. Put together, and taking into account dwarf-nova accelerated winds emanating from the central source, these data and models give a deeper insight into the GK Per nova remnant as a whole.
\end{abstract}
\runningtitle{Harvey \etal: GK Per Kinematics \dots}
\maketitle
\section{Introduction}
GK Per (1901) is a nearby, historic and spectacular source with a proximity of 470pc \cite{Harrison:2013aa}.
The system is one of only two classical novae observed within a planetary nebula to date,
offering a chance to study the evolution of both classes of object. 

A classical nova event is the result of thermonuclear runaway on the surface of a white dwarf accreting from, typically, a main sequence or a late G or K type 
star \cite{warner}. The accreted shell is ejected, at velocities that range from 5x$10^2$ to a few x$10^3$ km s$^{-1}$ 
\cite{BodeNova}. Dwarf novae, which are also exhibited by GK Per, are caused by an instability in the accretion disk surrounding the white dwarf star \cite{osakiDN}. These 
events accelerate winds to the order of 1-6x$10^3$ km s$^{-1}$, see \cite{dnvel2} and references therein. It is unknown if their ejection is spherically uniform or 
intrinsically bipolar \cite{lloydshaping,porterasphericity}. The common envelope phase is thought to play a major role in the 
shaping of nova remnants and planetary nebulae alike. Slower nova events are believed to have stronger deviations from 
spherical symmetry e.g. V1280 Sco \cite{Chesneau:2012aa} when compared to their more 
energetic counterparts. The system exhibits a pronounced stream of emission to the NE of the nova shell that is reminiscent of a jet but whose origin 
has not been settled. New and archival observations are used to explore the structure and kinematics of the GK Per nova shell.  

\section{Kinematics of the shell, knots and $\lq$jet'}
Using the Aristarchos telescope in Greece, new imaging was collected of GK Per in September 2014. 
The observations consisted of three narrow band filters (H$\alpha$, [N~{\sc ii}] and [O~{\sc iii}]), with 1800s exposures in each filter. 
Archival data of \cite{Liimets:2012aa} was used and of \cite{lawrencefp,Shara:2012aa}. The imaging was reduced using {\sc iraf}. To build a fuller view of the remnant and examine the $\lq$jet-like' feature, Manchester Echelle Spectrometer (MES) data were obtained at the San Pedro M\'artir observatory in Mexico. Position-Velocity (P-V) arrays were generated. The knot and shell P-V arrays were simulated using the morpho-kinematic modelling code {\sc shape}  \cite{shape}\footnote{A full description of {\sc shape} can be found at $http://bufadora.astrosen.unam.mx/shape/$}. A full discussion can be found in Harvey et al. (2016 in prep).


{\it Shell kinematics:} {\sc shape} modeling demonstrates that the shell most probably consists of an oblate elongated ring-like equatorial structure (PA {\raise.17ex\hbox{$\scriptstyle\sim$}} 120$^{\circ}$ and inc {\raise.17ex\hbox{$\scriptstyle\sim$}} 54$^{\circ}$) with polar over-densities (Fig. 1). 
The $\lq$boxy' nature of the nebula has long been observed e.g. \cite{Seaquist89}. The position angle derived of the old nova shell fits that of the $\lq$fossil' bipolar planetary nebula and the inclination to that of the binary. 
We find that the $\lq$polar caps' have a systematically lower expansion velocity than the rest of the shell. In the IRSA archives the longer wavelength WISE \cite{WISE} bands show
an abundance of material along these polar regions, potentially having a significant effect on their velocity structure evolution.

\begin{figure*}
\centering
\includegraphics[width=12.5cm]{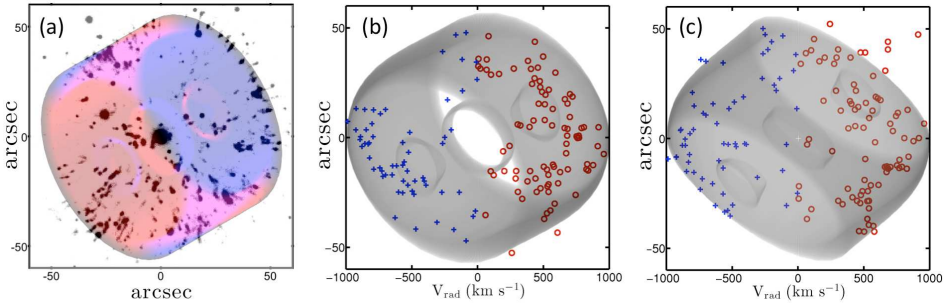}
\caption{Panel (a) shows the red-blue Doppler distribution as derived from P-V arrays as well as channel maps overlaid on an image from the Mayall telescope, north is up and east is to the left. Panel (b) shows the 
observed radial velocities of 148 knots dependant on their $\lq$y' position, i.e. from the north to south of the nebula, where the greyscale is the equivalent for the model created in  {\sc shape}. Panel (c) is the same as the middle panel except the positional arguments are from the east to west. }
\label{fig:explines}
\end{figure*}



\begin{figure}[!ht]
\centering
\includegraphics[width=12.5cm]{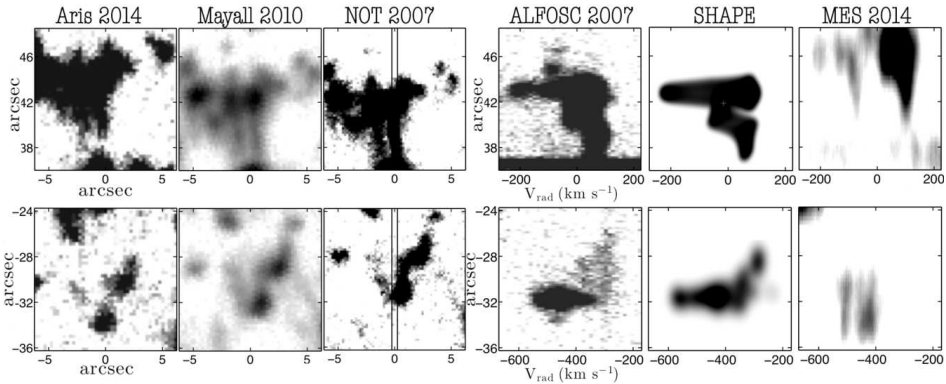}
\caption{Examples of knots progressing over time (2007-2014) and their corresponding P-V array and morpho-kinematic {\sc shape} model. Additional epoch P-V information is included, the knots here are along an axis of symmetry of the nova shell. The ALFOSC P-Vs correspond to the 2007 NOT image and the MES P-Vs correspond to the Aristarchos observations epoch.}
\label{fig:pvscomp}
\end{figure}

{\it Knot Characteristics:} In a clump-wind interaction their relative velocities must be considered, as it is believed to be the main mechanism in the shaping of the clumps. Under more uniform conditions, sub- sonic clumps have long tails and their supersonic counterparts display short-stubby tails \cite{Pittard:2005}. There are a variety of tail shapes present in GK Per suggesting diverse local flow conditions. The sinusoidal nature of the tails of the knots along the outer edge of the shell are evident in the 1997 Hubble images in \cite{Shara:2012aa}.
These wavy tails could be attributed to shaping by the dwarf nova winds, which can thus be found to 
have a velocity of {\raise.17ex\hbox{$\scriptstyle\sim$}} 4400 km s$^{-1}$. \cite{dngkper86} gave a velocity estimate of GK Per's dwarf nova winds 
of a few 1000 km s$^{-1}$. 

{\it Kinematics of the jet-like feature:} Previously there have been several theories to 
the origin of this feature, \cite{Bode04,Anupama:aa,Shara:2012aa}, first 
mentioned in \cite{anuprabhu}. Light-echo contours over IRAS imagery 
hinted that it predated the nova shell \cite{Bode04}. As our new kinematic data suggests a low 
velocity (\textless 17 km s$^{-1}$ with a Gaussian FHWM ranging from 30-40 km s$^{-1}$) we suggest 
that the feature may be an illuminated part of the waist of the fossil planetary nebula. 
This would also explain the curvature and spreading out of the feature. There is a 
slight augmentation in brightness of this line in the H$\alpha$ and {[O~{\sc iii}] 
5007~\AA} lines corresponding to the location of interaction with pre-existing material 
seen in radio and X-ray observations \cite{Anupama:aa,balogel,gkchandra15}.

\section{Discussion \& Questions}
One of the main findings from \cite{Liimets:2012aa} was there is no significant deceleration of the knots in the SW quadrant of the nova shell, though it had been long believed that the shell had been experiencing a stronger interaction with circumbinary material in this quadrant \cite{Seittermorph}. The barrel with polar over-densities 
structure presented here, as well as the basic bullet crushing time calculation support their finding. \cite{gkchandra15} indicates 
the system is moving through less dense material than during the time of the observations of \cite{balogel}. The derived morphology is decidedly oblate, in contrast to the shapes of other nova shells 
e.g. \cite{shapeRSoph,shape_munari_oph}. \cite{lloydshaping} used a 2.5D code to investigate remnant shaping for a variety of 
speed classes and produced rings, blobs and caps as expected but also created oblate remnants. Later \cite{porterasphericity} included the effects of 
a rotating accreted envelope, surprisingly the first panel in their Fig. 2 bares quite a resemblance to the morphology derived here. 


{\it SOKER:} Have you looked at the jets in NGC 40 that are perpendicular to the shell?

{\it HARVEY:} Interesting but no. I will investigate.

{\it HAMAGUCHI:} Does your model fit the X-ray morphology?

{\it HARVEY:} It suggests that the flattened shell is an intrinsic feature and is supported by the non-perpendicular proper motion of the system from 1917-1993, derived by Bode.

{\it DE MARCO:} On the neon overabundance, could GK Per be a born again ABG star?

{\it HARVEY:} In photoionisation simulations, we conducted using {\sc Cloudy}, a significant initial overabundance of neon was required. However, the stellar spectrum is not deficient in hydrogen nor enriched in helium, suggesting a nova origin is more likely.

\begin{acknowledgements}

E. Harvey wishes to acknowledge the support of the Irish Research Council for providing funding for this project under their postgraduate 
research scheme. The authors greatly benefitted from discussions with Dr. Wolfgang Steffen, Prof. John Meaburn, Dr. Myfanwy Llyod, Dr. Tomilsav Jurkic and Dr. Valerio Ribeiro, amongst others.

\end{acknowledgements}


\end{document}